\documentclass[%
 reprint,
 prx,
superscriptaddress,
 amsmath,amssymb,
 aps,
floatfix,
]{revtex4-2}

\usepackage{amsmath}
\usepackage{amsfonts}
\usepackage{amssymb}
\usepackage{hyperref}%
\usepackage[english]{babel}
\usepackage[T1]{fontenc}
\usepackage{lmodern}
\usepackage{svg}
\usepackage[utf8]{inputenc}
\usepackage[scaled=0.95]{helvet}
\usepackage{color}
\usepackage[normalem]{ulem}

\usepackage{xcolor,calc}
\hypersetup{linkcolor = black, citecolor = black, urlcolor = blue}

\providecommand{\U}[1]{\protect\rule{.1in}{.1in}}

\definecolor{darkgreen}{RGB}{2, 153, 0} 
\definecolor{darkyellow}{RGB}{204, 153, 0} 
\newenvironment{subfigure*}{}

\hypersetup{
colorlinks=true,
linkcolor=blue,
}

\usepackage{graphicx}
\usepackage{dcolumn}
\usepackage{bm}

\begin{document}

\preprint{APS/123-QED}

\title{Anatomy of spin-orbit-torque-assisted magnetization dynamics in Co/Pt bilayers: Importance of the orbital torque}
\author{Harshita Devda}

 \email{harshita.devda@uni-konstanz.de}
 \affiliation{Department of Physics, University of Konstanz, DE-78457 Konstanz, Germany}
\author{Andr\'{a}s De\'{a}k}
\affiliation{Department of Theoretical Physics, Institute of Physics, Budapest University of Technology and Economics, 1111 Budapest, Hungary}

\author{Leandro Salemi}
\affiliation{Department of Physics and Astronomy, Uppsala University, P. O. Box 516, S-751 20 Uppsala, Sweden}

\author{Levente R\'{o}zsa}
\affiliation{Department of Theoretical Solid State Physics, HUN-REN Wigner Research Centre for Physics, 1525 Budapest, Hungary}
\affiliation{Department of Theoretical Physics, Institute of Physics, Budapest University of Technology and Economics, 1111 Budapest, Hungary}

\author{László Szunyogh}
\affiliation{Department of Theoretical Physics, Institute of Physics, Budapest University of Technology and Economics, 1111 Budapest, Hungary}
\affiliation{HUN-REN-BME Condensed Matter Research Group, Budapest University of Technology and Economics, 1111 Budapest, Hungary}

\author{Peter M. Oppeneer}
\affiliation{Department of Physics and Astronomy, Uppsala University, P. O. Box 516, S-751 20 Uppsala, Sweden}

\author{Ulrich Nowak}
\affiliation{Fachbereich Physik, Universität Konstanz, DE-78457 Konstanz, Germany}

\date{\today}

\begin{abstract}

Understanding the mechanism driving magnetization switching in 
spin-orbit-torque-assisted devices remains a subject of debate. While originally attributed to the spin Hall effect and spin Rashba-Edelstein effect,
recent discoveries related to orbital moments induced by the orbital Hall effect and the orbital Rashba-Edelstein effect have added complexity to the comprehension of the switching process in non-magnet/ferromagnet bilayers. 
Addressing this challenge, we present a quantitative investigation  
of a Pt/Co bilayer by employing atomistic spin dynamics simulations, incorporating the proximity-induced moments of Pt, as well as electrically induced spin and orbital moments obtained from first-principles calculations. Our layer-resolved  model elucidates the damping-like and field-like nature of the induced moments by separating them according to their even and odd magnetization dependence. 
In addition to demonstrating {that} a larger {field-like} spin-orbit torque contribution {comes} from {previously disregarded} induced orbital moments, our work highlights the necessity of considering interactions with Pt induced moments at the interface, as they contribute significantly to the switching dynamics.

\end{abstract}

\maketitle

\section{Introduction}
Spin-orbit-torque (SOT)-based devices have emerged as promising candidates for non-volatile, high-speed, and energy-efficient storage applications \cite{Shao2021,Yi_Cao_2020,Xiufeng2021,Ji2024}.  Despite their potential, understanding the intricacies of controlled SOT-assisted switching in non-magnet/ferro-magnet (NM/FM) bilayers remains a persistent challenge. The complexity arises from the interplay of multiple phenomena that influence magnetization switching in specific directions differently  \cite{MihaiMiron2010,Miron2011,Liu.et.al,Hirsch1999, Gambardella2011, Edelstein1990, Dyakonov1971,Tanaka2008,Kontani2009,Dongwook2018,Salemi2019}.

Initially, SOT-induced switching was associated with two relativistic effects, the spin Hall effect (SHE) \cite{Hirsch1999,Sinova2015} and the spin Rashba-Edelstein effect (SREE) \cite{ Edelstein1990, Dyakonov1971, Haney2013a, Haney2013b, Freimuth2014, Emori2016, Moriya_2021}. However, recent studies have unveiled significant induced orbital moments \cite{Tanaka2008,Kontani2009,Dongwook2018,Salemi2019,Jo2024} alongside induced spin moments. 
These are attributed to the nonrelativistic orbital Hall effect (OHE) \cite{Tanaka2008,Kontani2009,Dongwook2018,Dongwook2020,Salemi2022OHE} and orbital  Rashba-Edelstein effect (OREE) \cite{Salemi2019}, whose occurrence introduces additional challenges to the understanding of switching dynamics. In addition, it has become evident that previously overlooked effects exist due to the magnetization of the FM layer, namely, the magnetic SHE (MSHE) \cite{Chuang2020}, magnetic SREE (MSREE), as well as the magnetic OHE and OREE (MOHE and MOREE, respectively) \cite{Salemi2022,salemi2021}. In contrast to the SHE, SREE, OHE and OREE that are time-reversal even effects, the magnetic counterparts are time-reversal odd and can induce an additional {self-torque} 
in the FM layer  \cite{salemi2021}. 
The origin of these distinct effects is intricate and involves various factors, including intrinsic band splitting, inversion-symmetry breaking, and interfacial spin-orbit coupling (SOC).

\begin{figure*}
    \includegraphics[width=1\textwidth]{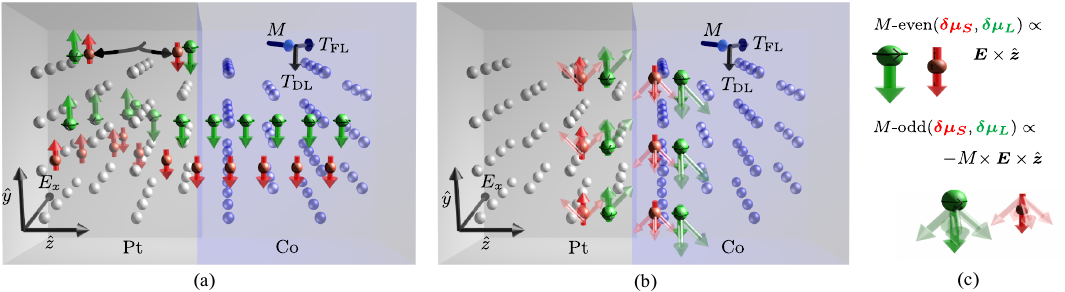}   
 \caption{Illustration of the various spin-orbit torque phenomena. (a) SHE and OHE generated spin (red) and orbital (green) moments in the non-magnet (Pt) diffuse into the ferromagnetic (Co) layers. (b) SREE and OREE generated spin and orbital moments induced at the Pt-Co interface. (c) All of the effects cause in addition intermixed $M$-even and $M$-odd induced moments. The direction of $M$-even induced moments is perpendicular to $\bm{E}$ and $\hat{\bm{z}}$, whereas $M$-odd induced moments vary with magnetization direction as  calculated from a Kubo linear-response formalism.}
 \label{fig:spin-orbit-phenomena-intro}
\end{figure*}

Upon applying an in-plane electric field in the $x$-direction, as depicted in Fig.~\ref{fig:spin-orbit-phenomena-intro}(a), the SHE results in a spin current perpendicular to the electric field.

At the interface, inversion-symmetry breaking in the presence of interfacial SOC leads to Rashba spin splitting in momentum space, giving rise to the relativistic  SREE 
causing induced spin moments, 
see Fig.~\ref{fig:spin-orbit-phenomena-intro}(b).

Beyond these spin-based effects, there is also the OHE \cite{Tanaka2008,Kontani2009,Dongwook2018,Salemi2022} 

giving rise to a perpendicular orbital current.

Similarly, the nonrelativistic OREE occurs at the inversion-symmetry breaking interface \cite{Salemi2019,salemi2021}, generating additional induced orbital moments.

Furthermore, there are also the magnetic effects mentioned above (MSHE, MOHE, and MSREE and MOREE) at the interface and within the interfaced conducting ferromagnetic layer, contributing to anomalous induced spin and orbital moments that are, e.g., perpendicular to those of the SHE and OHE \cite{salemi2021,Salemi2022,Kimata2019,Mook2020,Humphries2017,Abrao2025}. These induced moments apply  spin-orbit torque to the 
magnetization, see Fig.~\ref{fig:spin-orbit-phenomena-intro}(a) and (b), and can thus cause an anomalous self-torque \cite{Humphries2017,Bose2018b,Amin2018,Wang2019,TaoWang2020,Davidson2020,salemi2021}, manifesting as field-like and damping-like behavior and resulting in perpendicular magnetization switching.

Despite the advancements in understanding SOT, questions persist about its dynamics. Efforts to separate and comprehend the torques arising from specific effects have been made \cite{Haney2013a,Haney2013b,Freimuth2014, Wimmer2016, Amin2018,Belashchenko2019,Mahfouzi2018,Mahfouzi2020,Manchon2020,Hellman2017}.
Since the torque conserves the magnitude of the magnetic moment $\boldsymbol{M}$, it must be perpendicular to its direction. Torques of the form $\boldsymbol{T}_{\textrm{FL}}\propto\boldsymbol{M}\times\hat{\boldsymbol{P}}$ are called field-like, where the {unit vector} $\hat{\boldsymbol{P}}$ is conventionally associated with the polarization direction of the induced moments \cite{salemi2021}. Meanwhile, torques of the form $\boldsymbol{T}_{\textrm{DL}}\propto\boldsymbol{M} \times \left(\boldsymbol{M}\times\hat{\boldsymbol{P}}\right)$ are called damping-like. However, the intermixed nature of spin-orbit torque effects makes it challenging to link the underlying phenomena, specifically with damping-like and field-like behavior.
Moreover, SOTs are often discussed in terms of electrically generated relativistic spin polarization \cite{manchon2019,Hellman2017}, whereas the generated nonrelativistic orbital polarization has only recently been considered
\cite{Go2020,Lee2021,Go2020b,Go2023}.

First-principles calculations using the Kubo linear-response formalism reveal that the electrically induced moments are moreover not constant but vary with the changing magnetization direction over time \cite{salemi2021}. Writing the torque term generally as $\boldsymbol{T}\propto\boldsymbol{M}\times\delta \boldsymbol{\mu}_{S}$, it exhibits both field-like and damping-like behavior based on the components of the induced moment $\delta\boldsymbol{\mu}_{S}$ which are even and odd in the direction of the magnetization, respectively.

To unravel the complexity of SOT, it is hence imperative to consider {its} magnetization-dependent nature and systematically untangle the torque at the microscopic scale. 

With this motivation, our study aims to separate the torque behavior arising from electrically induced $M$-even and $M$-odd magnetization-dependent moments through a comprehensive analysis of the magnetization switching 
driven by them. 
The polarization direction of $M$-even and $M$-odd moments depicted in Fig.~\ref{fig:spin-orbit-phenomena-intro}(c) can be identified from the components of the magneto-electric susceptibility tensor calculated from first principles. 
Additionally, our model elucidates the effects arising from induced spin and orbital moments separately. 
Furthermore, our 
description addresses the challenge of incorporating interactions with proximity-induced moments in the non-magnet within a purely magnetic model through renormalization, offering {thus} a microscopic perspective on the intricate interplay of various phenomena and interactions.

\section{Methodology}

\subsection{Spin model in the absence of electric field}

In our model for the Co/Pt bilayer, we consider strong spin magnetic moments, $\boldsymbol{\mu}_{S,i}=\mu^d_{S,i} \boldsymbol{S}_{i}$ with $|\boldsymbol{S}_{i}|=1$ for the cobalt atoms, where the magnitudes of the moments $\mu^d_{S,i}$ related to the $3d$ electrons are supposed to be stable against changing their  orientation 
$\{ \boldsymbol{S}_{i} \}$. On the contrary, proximity with the Co moments induces weak spin magnetic moments $\boldsymbol{\mu}_{S,\nu}$ at the Pt atoms that vary both in magnitude and direction depending on the configuration of the Co spins. Including 
spin-orbit effects this model can well be described in terms of an extended Heisenberg spin Hamiltonian,   
\begin{align}
   \mathcal{H}(\{\boldsymbol{S}_{i}\},\{\boldsymbol{\mu}_{S,\nu}\}) &=  \sum_{i}\boldsymbol{S}%
_{i}\mathcal{K}_{i}\boldsymbol{S}_{i}-\frac{1}{2}\sum_{i\neq j}\boldsymbol{S}%
_{i}\mathcal{J}_{ij}\boldsymbol{S}_{j} \nonumber \\
& ~~ -\sum_{i\nu}\boldsymbol{S}%
_{i}\mathcal{L}_{i\nu}\boldsymbol{\mu}_{S,\nu}+\sum_{\nu} \mathcal{H}_{\nu}%
(\boldsymbol{\mu}_{S,\nu})\,, \label{eq:Ham-zero-field}%
\end{align}
where  
$\mathcal{K}_{i}$ represents the symmetric on-site anisotropy matrix, while $\mathcal{J}_{ij}$ and $\mathcal{L}_{i\nu}$ stand for the tensorial exchange couplings 
between the Co atoms and between the Co and Pt atoms, respectively. 
In particular, the exchange interaction between the Heisenberg-type Co spins can be separated into three contributions,
\begin{equation}
     \boldsymbol{S}_i {\mathcal{J}}_{ij}  \boldsymbol{S}_j = {J}_{ij} \boldsymbol{S}_i \cdot \boldsymbol{S}_j + \boldsymbol{S}_i \mathcal{J}_{ij}^S  \boldsymbol{S}_j + \boldsymbol{D}_{ij} (\boldsymbol{S}_i \times \boldsymbol{S}_j) \, ,\label{eq:exc_tensor}%
\end{equation}
with the isotropic exchange interaction strength $J_{ij}=\frac{1}{3} {\rm Tr} {\mathcal{J}}_{ij}$ \cite{LKAG1987} and the traceless symmetric exchange anisotropy matrix $\mathcal{J}_{ij}^S=\frac{1}{2}({\mathcal{J}}_{ij}+{\mathcal{J}}^T_{ij})- J_{ij} {\mathcal{I}}$, with ${\mathcal{I}}$ being the unit matrix. The third term on the right-hand side 
of Eq.~\eqref{eq:exc_tensor} 
describes the Dzyaloshinskii–Moriya interaction (DMI)~\cite{DZYALOSHINSKY1958,Moriya1960} with the DM vector $\boldsymbol{D}_{ij}$, which is associated with the three independent elements of the asymmetric part of the exchange tensor, $\frac{1}{2}({\mathcal{J}}_{ij}-{\mathcal{J}}^{T}_{ij})$. 

In the above model, the size of the induced Stoner-type spin moments of the Pt atoms is not fixed, but it is determined by the balance between the Co-Pt exchange interaction and the formation energy of the Pt spin moments $\mathcal{H}_{\nu}(\boldsymbol{\mu}_{S,\nu})$. For the latter  we take the simplest approximation,%
\begin{equation}
\mathcal{H}_{\nu}(\boldsymbol{\mu}_{S,\nu})
=a_{\nu} \mu_{S,\nu}^{2}%
\,\quad\left(  a_{\nu}>0\right) ,\label{eq:creation_energy}
\end{equation}
where $a_{\nu}$ is a constant to be determined from \textit{ab initio} calculations. Assuming that the formation of the induced Pt moments occurs on a much shorter time scale than the motion of the strong Co spin moments, the former ones can be eliminated from the spin Hamiltonian \eqref{eq:Ham-zero-field} following the methods proposed by Mryasov \cite{Mryasov2005} and Le\ifmmode \check{z}\else \v{z}\fi{}ai\ifmmode \acute{c}\else \'{c}\fi{} {\em et al.} \cite{Lezaic2013}. The 
renormalized spin Hamiltonian then reads
\begin{equation}
     \mathcal{H}(\{\boldsymbol{S}_{i}\})=  \sum_{i} \boldsymbol{S}_{i} {\widetilde{\mathcal{K}}_{i}} \boldsymbol{S}_{i} -\frac{1}{2}\sum_{i\neq j} \boldsymbol{S}_{i}\widetilde{\mathcal{J}}_{ij}\boldsymbol{S}_{j} \, ,  \label{eq:Ham_renorm-zero-field} 
\end{equation}
with the renormalized Co-Co exchange tensor 
\begin{equation}
{\widetilde{\mathcal{J}}_{ij}}=\mathcal{J}_{ij}+\sum_{\nu}%
\frac{\mathcal{J}_{i\nu}\mathcal{J}_{\nu j}}{|\sum_{n}\boldsymbol{S}%
_{0}\mathcal{J}_{\nu n}\boldsymbol{S}_{0}|} \label{eq:J_renorm}
\end{equation}
and the renormalized on-site anisotropy tensor 
\begin{equation}
{\widetilde{\mathcal{K}}_{i}}= \mathcal{K}_{i}-\frac{1}{2}\sum_{\nu
}\frac{\mathcal{J}_{i\nu}\mathcal{J}_{\nu i}}{|\sum_{n}\boldsymbol{S}%
_{0}\mathcal{J}_{\nu n}\boldsymbol{S}_{0}|} \, . \label{eq:K-renorm}
\end{equation}
In the above expressions, $\boldsymbol{S}_{0}$ denotes the orientation of the magnetization in the ferromagnetic 
state and $\mathcal{J}_{i\nu}=\mathcal{L}_{i\nu}/|\mu_{S,\nu}^{\rm FM}|$, where $\mu_{S,\nu}^{\rm FM}$ is the induced moment at the Pt site $\nu$ in the FM state. Note that we assumed that the size of the Weiss field at the Pt sites,  $h_\nu = |\sum_{n}\boldsymbol{S}_{0}\mathcal{J}_{\nu n}\boldsymbol{S}_{0}|$ 
and also $\mu_{S,\nu}^{\rm FM}$ weakly depend on $\boldsymbol{S}_{0}$. 
Our numerical results justify this assumption. The details of the derivation can be found in  
Sec. I of the Supplemental Material  \cite{SupMat}.

\subsection{Theory for electrically induced moments
\label{sec:IIB}}

We employ the Kubo linear-response formalism  to compute \textit{ab initio} the electric-field response of the NM/FM system. There are generally two magnetic quantities that appear in response to an applied electric field, the spin and orbital Hall conductivities and the Rashba-Edelstein or magneto-electric (ME) susceptibility,
\begin{align}
    \bm{j}_{S/L}= \bm{\sigma}^{S/L}\cdot\bm{E} ~~ \textrm{and} ~~~ \delta\bm{\mu}_{S/L} = \bm{\chi}^{S/L} \cdot \bm{E},
\end{align}for spin ($S$) and orbital ($L$) angular momentum, respectively. In the situation considered here, for a finite bilayer system and for a stationary electric field, the SHE and OHE conductivities will lead to an accumulation of spin and orbital moments at the boundaries of the NM/FM bilayer.
The electrically induced spin and orbital magnetic moments can then locally be described by
\begin{align}
\delta \boldsymbol{\mu}_{S/L,i/\nu} & = \bm{\chi}^{S/L}_{i/\nu}(\theta,\phi) \cdot \boldsymbol{E} \, \label{eq:ind_mu_propto_E}, 
\end{align} 
where the indices denote the Co ($i$) and Pt ($\nu$) sites. The ME tensor and, hence, the  induced moments 
depend on the {direction of the FM} magnetization {defined by} angles $\theta$ and $\phi$,
with the $\hat{\bm{z}}$ axis  perpendicular to the NM/FM interface and $\theta=90^{\circ}$ and $ \phi =0^{\circ}$ corresponding to the in-plane $\hat{\bm{x}}$ direction, along which the electric field is applied,   see Fig.\ \ref{fig:spin-orbit-phenomena-intro}(a). 
For simplicity, we rewrite $\bm{\chi}^{S/L}_{i/\nu}$ as $\left[\bm{\chi}\right]^{\eta}$ where $\left[ ...\space \right]^{\eta}$ can take four values  $\left[ ... \right]^{S}_{i}$, $\left[ ... \right]^{S}_{\nu}$,$\left[... \right]^{L}_{i}$and $\left[... \right]^{L}_{\nu}$. The index $i$ runs over the eight Co layers and $\nu$ over the eight Pt layers of the considered Pt/Co bilayer (see Section \ref{sec:IV}  below).

The angle dependence of the ME susceptibility tensors has been analyzed before  \cite{salemi2021}. Up to the second-order terms in magnetization direction, it is rigorously given as (see Sec. II in the Supplemental Material \cite{SupMat})
\begin{widetext} 
\begin{equation}
\left[\bm{\chi}\right]^{\eta}(\theta,\phi)= \begin{bmatrix} [C_{1}]^{\eta}\cos\theta + [C_{2}]^{\eta}(1-\cos2\theta) \sin2\phi  &  [C_{3}]^{\eta} -[C_{2}]^{\eta} [1-\cos2\theta]\cos2 \phi  & [C_{5}]^{\eta}\sin \theta \cos \phi \\
& - [C_{4}]^{\eta}\cos2\theta & +[C_{6}]^{\eta}\sin2\theta \sin\phi  \\
& & &\\
	  -[C_{3}]^{\eta} -[C_{2}]^{\eta} [1-\cos2\theta]\cos2 \phi  &  [C_{1}]^{\eta}\cos\theta -[C_{2}]^{\eta}(1-\cos2\theta) \sin2\phi & [C_{5}]^{\eta}\sin \theta \sin \phi  \\
	 + [C_{4}]^{\eta}\cos2\theta &   &  - [C_{6}]^{\eta}\sin2\theta \cos\phi\\
	  & & &\\
	[C_{7}]^{\eta}\sin \theta \cos \phi + [C_{8}]^{\eta}\sin2\theta \sin\phi &[C_{7}]^{\eta}\sin \theta \sin \phi - [C_{8}]^{\eta}\sin2\theta \cos\phi & [C_{9}]^{\eta}\cos\theta \\ 
	\end{bmatrix} . \label{eq:ME_tensor}
\end{equation} 
\end{widetext}

Here, the coefficients $[C_1]^{\eta} \cdots [C_9]^{\eta}$ describe the complete $3 \times 3$ tensor, for arbitrary directions of the electric field. Each coefficient depends implicitly on the layer index ($i$ or $\nu$) and on $\eta = S$ or $L$. The tensor elements include time-reversal even and odd contributions from SHE, OHE,  SREE and OREE, as well as their magnetic counterparts. As a result,
the components of $[\bm{\chi}]^{\eta}(\theta,\phi)$ are either invariant or change their sign upon reversal of the magnetization, termed as $M$-even or $M$-odd-like dependence. This $M$-even or $M$-odd dependence will affect the electric field-induced spin and orbital moments correspondingly. 
The $M$-even  components of the ME susceptibility tensor $[\bm{\chi}]^{\eta}(\theta,\phi)$ 

are then given as 
\begin{widetext} 
\begin{equation}
[\bm{\chi}]^{\eta}_{\text{even}}(\theta,\phi)= \begin{bmatrix}  [C_{2}]^{\eta}(1-\cos2\theta) \sin2\phi  &  [C_{3}]^{\eta} -[C_{2}]^{\eta} [1-\cos2\theta]\cos2 \phi  &  [C_{6}]^{\eta}\sin2\theta \sin\phi\\
& - [C_{4}]^{\eta}\cos2\theta &   \\
& & &\\
	  -[C_{3}]^{\eta} -[C_{2}]^{\eta} [1-\cos2\theta]\cos2 \phi  &   -[C_{2}]^{\eta}(1-\cos2\theta) \sin2\phi &  - [C_{6}]^{\eta}\sin2\theta \cos\phi \\
	 + [C_{4}]^{\eta}\cos2\theta &   & \\
	  & & &\\
	 [C_{8}]^{\eta}\sin2\theta \sin\phi & - [C_{8}]^{\eta}\sin2\theta \cos\phi & 0 \\ 
	\end{bmatrix} , \label{eq:ME_even_tensor}
\end{equation} 

and the $M$-odd components as,

\begin{equation}
[\bm{\chi}]^{\eta}_{\text{odd}}(\theta,\phi)= \begin{bmatrix} [C_{1}]^{\eta}\cos\theta   &  0  & [C_{5}]^{\eta}\sin \theta \cos \phi \\
& & &\\
	  0  &  [C_{1}]^{\eta}\cos\theta  & [C_{5}]^{\eta}\sin \theta \sin \phi \\
	  & & &\\
	[C_{7}]^{\eta}\sin \theta \cos \phi &[C_{7}]^{\eta}\sin \theta \sin \phi  & [C_{9}]^{\eta}\cos\theta \\ 
	\end{bmatrix}  .\label{eq:ME_odd_tensor}
\end{equation} 
\end{widetext}
The coefficients $[C_j]^{\eta}$ can be deduced from the tensor elements as explained in Sec.~\ref{sec:IV}. Furthermore, for an electric field along the $\hat{\bm{x}}$ direction, the case we will consider in the following, only the first column of the tensor will contribute to the SOT.

\subsection{Spin model in the presence of electrically induced moments\label{sec:III}}

To implement the interactions from electrically-induced moments along with the proximity-induced Pt moments, we modify the Heisenberg Hamiltonian of Eq.~\eqref{eq:Ham-zero-field}, as explained in Sec. I of the Supplemental Material \cite{SupMat},
\begin{align}
   & \mathcal{H}(\{ \boldsymbol{S}_i \}, \{ \boldsymbol{\mu}_{S,\nu} \}, \boldsymbol{E}) = \sum_{i}  \boldsymbol{S}_{i} \mathcal{K}_{i}\boldsymbol{S}_{i} \nonumber \\
   & -\frac{1}{2}\sum_{i\neq j} (\boldsymbol{S}_{i}+\frac{\delta\bm{\mu}_{S,i}}{\mu_{S}^{d}}) 
   {\mathcal{J}}_{ij} (\boldsymbol{S}_{j}+\frac{\delta\bm{\mu}_{S,j}}{\mu_{S,}^{d}}) 
    -\! \sum_{i} J^{sd} \, \frac{\delta\bm{\mu}_{S,i}}{\mu_{S}^{d}} \cdot \boldsymbol{S}_{i} \nonumber \\ 
     &  -\! \sum_{i\nu} (\boldsymbol{S}_{i}+\frac{\delta\bm{\mu}_{S,i}}{\mu_{S}^{d}}) \mathcal{L}_{i\nu} \boldsymbol{\mu}_{S,\nu} 
     - \! \sum_{\nu} \boldsymbol{b}^{(d)E}_{\nu}\boldsymbol{\mu}_{S,\nu} + \! \sum_\nu \mathcal{H}_{\nu}(\boldsymbol{\mu}_{S,\nu}) \nonumber \\ 
     & + \!\sum_{i}\frac{\zeta_{i}}{2\mu_{B}^{2}} \, \mu^{d}_{S,i}  \boldsymbol{S}_{i} \cdot {\delta\bm{\mu}_{L,i}} 
      +\! \sum_{\nu}\frac{\zeta_{\nu}}{2\mu_{B}^{2}}\boldsymbol{\mu}_{S,\nu} \cdot {\delta\bm{\mu}_{L,\nu}} \, . 
      \label{eq:ham-with-field}
\end{align}
Here, the exchange interaction of the electrically-induced spin moments of the Co atoms with Co and Pt spin moments is included in the second and fourth terms, respectively, while the on-site exchange interaction between the strong ($d$ electron) spin moments and the electrically induced ($s$ electron) spin moment is incorporated in the third term as an isotropic intra-atomic exchange with $J^{sd} =3.6 ~\textrm{mRy}$.  
 
Although this value of the intraatomic exchange is only an estimate, our calculations confirmed that considering values of $J^{sd}$ up to 36 mRy does not change the main conclusions about the relative strength of the spin and orbital contributions to the torque.
Since the spin magnetic moment at the Pt sites induced by the electric field, $\delta \boldsymbol{\mu}_{S,\nu} = [\bm{\chi}]^S_\nu \cdot \boldsymbol{E}$, contains a contribution from the exchange interaction with the electrically-induced Co moments included in the fourth term, the contribution related to a direct spin-polarization effect at the Pt sites is taken into account through an effective field, $\boldsymbol{b}_{\nu}^{(d)E}$, in the fifth term of Eq.~\eqref{eq:ham-with-field}. Finally, the last two terms represent the spin-orbit coupling interaction between  spin moments and electrically-induced orbital moments at Co and Pt sites. The semiclassical treatment of the spin-orbit coupling term is explained in Sec. III of the Supplemental Material
\cite{SupMat}. Here, $\zeta_{i}$ and $\zeta_{\nu}$ stand for the spin-orbit coupling strength at the Co and Pt atoms, respectively. These are  for Co $\zeta_{\textrm{Co}}=6.67~\textrm{mRy}$ and for Pt $\zeta_{\textrm{Pt}}=44.20~\textrm{mRy}$, as estimated from our first-principles calculations.
Similar to the procedure leading to Eq.~\eqref{eq:Ham_renorm-zero-field}, the Hamiltonian in Eq.~\eqref{eq:ham-with-field} can be renormalized, removing the induced Pt spin moments, as detailed in the Sec. I of the Supplemental Material \cite{SupMat}. We end up with the following spin Hamiltonian,
\begin{align}
   & \mathcal{H}(\{\boldsymbol{S}_i\}, \boldsymbol{E})=  
   \sum_{i} \boldsymbol{S}_{i} {{\widetilde{\mathcal{K}}_{i}}} \boldsymbol{S}_{i} -\frac{1}{2}\sum_{i\neq j} \boldsymbol{S}_{i}{\widetilde{\mathcal{J}}_{ij}} \boldsymbol{S}_{j}  
     \nonumber \\
   & ~~  -\sum_{i\neq j} \boldsymbol{S}_{i} \mathcal{J}_{ij} \frac{\delta\bm{\mu}_{S,j}}{\mu_{S,i}^{d}}-\sum_{i} J^{sd} \, \frac{\delta\bm{\mu}_{S,i}}{\mu_{S,i}^{d}} \cdot \boldsymbol{S}_{i} 
   \nonumber \\
    & ~~ - \sum_{i,\nu}  \boldsymbol{S}_{i} \frac{\mathcal{J}_{\nu i}}{|\mu_{\nu}^{\mathrm{FM}}|} \delta\bm{\mu}_{S,\nu}
      + \sum_{i} \frac{\zeta_{i}\mu^d_{S,i} }{2\mu_{B}^{2}}  {\delta\bm{\mu}_{L,i}} \cdot
    \boldsymbol{S}_{i} 
    \nonumber \\
    & ~~ +  \sum_{i,\nu} \frac{\zeta_{\nu} |\mu_{\nu}^{\mathrm{FM}}| }{2\mu_{B}^{2}  |\sum
    _{n}\boldsymbol{S}_{0} \mathcal{J}_{\nu n} \boldsymbol{S}_{0}|}   {\delta\bm{\mu}_{L,\nu}} \mathcal{J}_{\nu i} \boldsymbol{S}_{i}  \nonumber \\
    &~~ = \mathcal{H}_{\rm{Ani}}+\mathcal{H}_{\rm{Exc}}+\mathcal{H}_{\rm{Co}}^{S}+\mathcal{H}_{\rm{Intra}}^{S}+ \mathcal{H}_{\rm{Pt}}^{S}+\mathcal{H}_{\rm{Co}}^{L}+\mathcal{H}_{\rm{Pt}}^{L}\, ,
    \label{eq:ham-renorm-with-field}
    \end{align}
to be used in our spin dynamics simulations. Here, quadratic terms in the electric-field-induced quantities have been neglected.
The electrically-induced spin and orbital moments calculated from the ME susceptibility tensor (Eq.~\eqref{eq:ME_tensor}) are implemented layerwise.
As the magnetization direction changes with time during the switching process, the ME tensor $[\bm{\chi}]^{\eta} (\theta,\phi)$ is dynamically modified which is fully included through its  time-varying angle dependence.

\subsection{Details of the {\em ab initio} calculations
\label{sec:IV}}

First, the atomic positions of the bilayer system consisting of eight atomic monolayers of Co on eight layers of Pt(001) have been fully relaxed with the Density Functional Theory (DFT) package SIESTA \cite{SIESTA}. 
For the in-plane lattice constant of the fcc(001) planes we obtained $a_{\mathrm {2D}}=2.763 \, \AA$, while the interlayer distance of the interior Pt layers was $d_0=2.0338 \, \AA$. Note that this value is slightly larger than the ideal interlayer distance of the fcc(001) planes $d_{\rm 3D}=\frac{\sqrt{2}}{2}a_{\rm 2D}=1.9537 \, \AA$. For the Pt layers near the interface and for the Co layers we found a remarkable inward relaxation as listed in Table~\ref{tab:relaxations}.

\begin{center}
\begin{table*}[tbh!]
    \caption{Layer geometry data from  DFT (SIESTA) calculations along with the spin magnetic moments from  the SKKR method. $d$ is  the distance of the atomic layer specified in the first row to the previous atomic layer. $r=(d-d_0)/d_0$ quantifies the relaxation of the layers,  where $d_0=2.0338 \, \AA$ is the distance between layers Pt$_5$ and Pt$_4$.
     \label{tab:relaxations}}
    \begin{tabular}{c|ccccccccccc}
    \hline\\[-0.2cm]
        layer   & Pt$_3$ & Pt$_2$ & Pt$_1$ & Co$_1$ & Co$_2$ & Co$_3$ & Co$_4$ & Co$_5$ & Co$_6$ & Co$_7$ & Co$_8$  \\ [6pt] \hline  \\ [-6pt]
        $d$ (\AA) & 2.0330 & 2.0435 & 2.0659 & 1.7756 & 1.4909 & 1.4837 & 1.4670 & 1.4730 & 1.4471 & 1.5178 & 1.4496  \\ [6pt]
        $r$ (\%)  & -0.04   & 0.48  & 1.58  & -12.69  & -26.69 & -28.05 & -27.87 & -27.58 & -28.85 & -25.37 & -33.64 \\ [6pt]
        $\mu_{\rm spin}$ ($\mu_{\rm B}$) 
                  & 0.016   & 0.063 & 0.206 &  1.88   &  1.78  &  1.73  &  1.72  &  1.74  &  1.70  &  1.66  &   1.93
                  \\[0.2cm]
                  \hline

    \end{tabular}
\end{table*}
\end{center}

Second, to determine the electronic structure and magnetic interactions of the Co/Pt(001) bilayer, we used the fully relativistic Screened Korringa-Kohn-Rostoker (SKKR) method \cite{zabloudil_electron_2005}. Here the Pt host is treated as a semi-infinite bulk system in the ideal fcc geometry with the lattice constant $a_{\rm 3D}=3.9075 \, \AA$ corresponding to $a_{\rm 2D}$ and $d_{\rm 3D}$ as given above. On top of the semi-infinite Pt bulk host eight atomic layers of Pt and eight atomic layers of Co are deposited in a geometry approximating what has been obtained with the SIESTA code: the small relaxation of the Pt layers was neglected, the positions of the Co layers at the interface (Co$_1$), of the interior Co layers (Co$_2$ to Co$_7$) and of the surface Co layer (Co$_8$) were relaxed by -13\%, -28\% and -34\%, respectively. 
In the SKKR calculations we used the Local Spin-Density Approximation (LSDA) \cite{VWN-LDA-1980}, 
the potentials were treated within the atomic-sphere approximation (ASA) with an angular momentum cutoff of $\ell_\text{max}=2$.
In order to relax the potentials toward the vacuum represented by a constant potential, four layers of empty spheres have been applied. 

As shown in Table~\ref{tab:relaxations}, the spin magnetic moments of the Co layers inside the Co film are in the range of $1.70-1.78$ $\mu_{\rm B}$ being close to the bulk Co value, while at the interface and surface they show a remarkable enhancement. Note also the induced spin magnetic moment of the Pt layers, particularly at the interface.

Fixing the effective potentials obtained from the self-consistent field (scf) SKKR calculations, we employed the Relativistic Torque Method (RTM) \cite{Udvardi2003} in the spirit of the magnetic force theorem to compute the tensorial exchange interactions $\mathcal{J}_{ij}$ and $\mathcal{J}_{i\nu}$, as well as the on-site anisotropy matrices $\mathcal{K}_i$ in the spin model \eqref{eq:Ham-zero-field}. 
Due to the $C_{4v}$ symmetry of the system, for each layer the on-site anisotropy energy can simply be written as $K_i (S_i^z)^2$, where $K_i$ is the uniaxial anisotropy constant. The exchange interactions have been calculated within a sphere of radius of $3a_{\rm 2D}$ around a representative site in each layer.  

The isotropic interactions between the Co atoms are strongly ferromagnetic: the nearest-neighbor (NN) out-of-plane interactions range between 1.5 to 4 mRy, the NN in-plane interactions between 0.3 to 2 mRy, while the interactions for the farther Co-Co pairs fall off quickly with the distance. Importantly, the renormalized isotropic interactions $\tilde{J}_{ij}$ differ only a little from the raw interactions $J_{ij}$. The DM interactions are of negligible importance in this system, as the length of all DM vectors is below 0.1 mRy. From the renormalized spin Hamiltonian we calculated a total magneto-crystalline anisotropy (MCA) energy per 2D unit cell, $E[\bm{M}\,||\,\hat{\bm{z}}] - E [\bm{M}\,||\,\hat{\bm{x}}] =0.234 \, {\rm mRy}$. This implies an in-plane ground-state magnetization for our bilayer, see 
Sec. IV in the Supplemental Material~\cite{SupMat} for details.

Inferring the layer-resolved MCA energies, the Co layer at the Co/Pt interface (Co$_1$) has a negative (out-of-plane) contribution of $-0.018$~mRy, the contributions of the interior Co layers (Co$_2-$Co$_7$) scatter between $0.025-0.042$~mRy, while the surface Co layer (Co$_8$) adds the largest 
in-plane contribution of nearly 0.060~mRy.

Lastly, the electrically induced spin and orbital moments were calculated with the relativistic DFT package WIEN2k 
 \cite{blaha2019,Kunes2001},
using the implementation of Ref.~\cite{salemi2021}, using the constant relaxation time approximation ($\hbar \tau^{-1} = 0.15$ eV).
Specifically, we calculated the layer-dependent spin and orbital ME tensors for 13 different orientations of the Co layer magnetization. We then use the exact trigonometric dependence of the $[\bm{\chi}]^{\eta} (\theta, \phi)$ (Eq.~(\ref{eq:ME_tensor})) to fit the coefficients $[C_j]^{\eta}$ for all Co and Pt layers.  Using these fitting coefficients and  calculating analytically the induced moments dependence on the average spin direction ($\boldsymbol{S}$), 

electric field vector ($\boldsymbol{E}$) and vector normal to interface plane ($\hat{\boldsymbol{z}}$) \cite{SupMat}, we see that the $M$-even induced moments have the general form, 
\begin{align}
   \delta \boldsymbol{\mu}^{e}_{S/L}& = [P(A)]^{\eta}(\boldsymbol{E} \times \hat{\boldsymbol{z}}) \nonumber \\
    & +[ P(A')]^{\eta}(\boldsymbol{S}  \cdot \boldsymbol{E} )(\boldsymbol{S} \times \hat{\boldsymbol{z}})\nonumber \\
    & +[P(A'')]^{\eta} \boldsymbol{S}\left[ \boldsymbol{E} \cdot (\boldsymbol{S} \times \hat{\boldsymbol{z}} )\right]  \nonumber \\ 
    & + [P(A''') ]^{\eta}(\boldsymbol{E} \times \boldsymbol{S})(\boldsymbol{S} \cdot \hat{\boldsymbol{z}}) ,
    \label{eq:M-even_directions}
\end{align}
while the $M$-odd induced moments have the form,
\begin{equation}
    \delta \boldsymbol{\mu}^{o}_{S/L} =-[P(B)]^{\eta} \boldsymbol{S} \times (\boldsymbol{E} \times \hat{\boldsymbol{z}}) + [P(B')]^{\eta}(\boldsymbol{S} \cdot \boldsymbol{E}) \hat{\boldsymbol{z}} ,
    \label{eq:M-odd-directions}
\end{equation}
which will further on 
be split in contributions to the 

field-like and  damping-like torques, in agreement with previous works  \cite{Manchon2020, Liu2021,Liu2022,Xue2023}.  For the electrically induced moments in the Co layers $\boldsymbol{S}$ represents the average spin direction of the total Co system, whereas for the induced moments in Pt, we have considered $\boldsymbol{S}$ as average spin direction of the interface Co layer.  In case of an in-plane magnetization with $\boldsymbol{S} \, || \, \boldsymbol{E}$, a planar field-like term will only contain  coefficient $P(A')$ and a planar damping-like term $P(B')$ only. { These coefficients are related to each other and to the coefficients $[C]^\eta$  of Eq. \eqref{eq:ME_tensor} as  follows: $P(A)=$ $(C_{3}-2C_{2}+C_{4})/2, P(A'')=t_{m} P(A), P(A')=4C_{2}-P(A''),P(A''')=2C_{8}-P(A''), P(B)=-C_{1}$ and $P(B')=C_{7}-C_{1}$. Here, it is not possible to find the exact separation of $P(A) $ and $ P(A'')$, as both  act in combination for $\boldsymbol{S}\, || \, (\boldsymbol{E} \times \hat{\boldsymbol{z}})$. We assume that for $\boldsymbol{S} \, || \,(\boldsymbol{E} \times \hat{\boldsymbol{z}})$ the net induced moments will be large and pointing in $\boldsymbol{E} \times \hat{\boldsymbol{z}}$ direction  and, hence, we choose $t_m=-1$ for the further analysis. We will later show that our assumption matches perfectly with the obtained changes in spin and orbital induced moments from the magnetization dynamics, where we have implemented the induced moments in form of the ME tensor given in Eq.\ \eqref{eq:ME_tensor}  with coefficients $[C]^\eta$. The analytical form in Eqs.\ (\ref{eq:M-even_directions}) and (\ref{eq:M-odd-directions})  are utilized for a clearer understanding only.}

\begin{figure*}
    \includegraphics[scale=0.9]{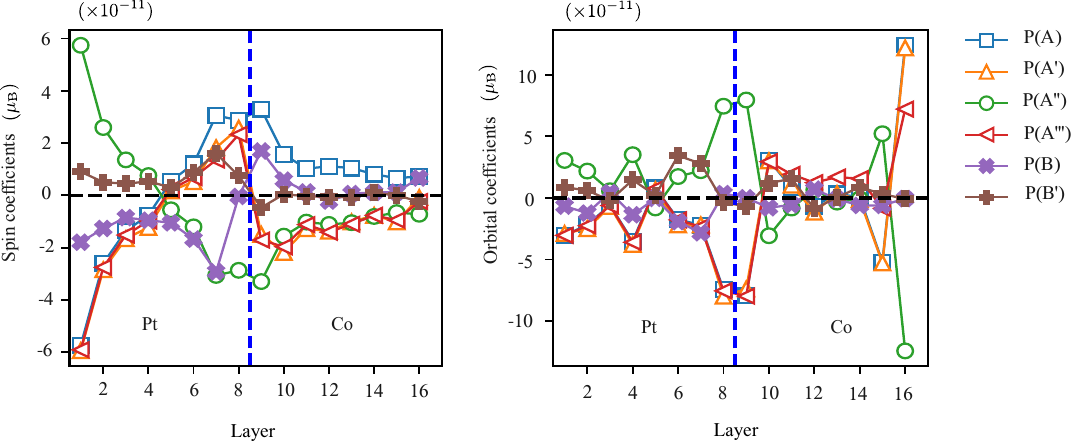}
    \caption{\textit{Ab initio} calculated expansion coefficients $P(A), P(A'), P(A''), P(A''')$ and $P(B)$, $P(B')$ of the induced spin (left) and orbital (right) moments according to Eqs.~\eqref{eq:M-even_directions} and \eqref{eq:M-odd-directions} in a Pt/Co bilayer. 
    The layer-resolved moments are plotted for 8 Pt  layers (indices $1-8$) and 8 Co layers (indices $9-16$).}
    \label{fig:MES_constants}
\end{figure*}

The dominating term in  Eqs.~\eqref{eq:M-even_directions} and \eqref{eq:M-odd-directions} can be determined by comparing the magnitudes of the coefficients of each term as shown in Fig.~\ref{fig:MES_constants} for an in-plane electric field. This leads to the following 
observations regarding the different 
origins of the induced moments:
\begin{itemize}
    \item $P(A)$ describes the $M$-independent (and thus $M$-even) induced spin/orbital moments due to the time-reversal even SHE and OHE and the SREE and OREE.  
    \item Terms with coefficients $P(A')$, $P(A'')$, and $P(A''')$ are {magnetization dependent, but} time-reversal even. Magnetization-dependent terms arise due to spin scattering and spin-filtering at the interface. This effect can be interpreted as a result of interfacial symmetry breaking. 
    \item Due to the MSHE, MOHE, as well as MSREE and MOREE 
    there will be additional induced moments, represented by terms with coefficients $P(B)$ and $P(B')$, which are $M$-odd \cite{salemi2021,Salemi2022}. 
    For the induced spin moments, these elements depend on SOC and are,  hence, larger on the Pt side of the interface. For the orbital moments, these elements are very small. 
\end{itemize}

Further below, we will see in detail how these terms separately affect the magnetization dynamics via their different contributions to the SOT.

\subsection{Atomistic spin dynamics simulations}

Spin dynamics at finite temperature is  described by the stochastic Landau-Lifshitz-Gilbert (SLLG) equation~\cite{Nowak2007},
\begin{equation}
                 \boldsymbol{\dot{S}}_{i} = -\frac{\gamma}{(1+\alpha^{2}) \mu_{S,i}^{d}} \boldsymbol{S}_{i} \times [\boldsymbol{H}_{i}^{\rm eff}  + \alpha \boldsymbol{S}_{i} \times \boldsymbol{H}_{i}^{\rm eff}] ,\label{eq:S-LLG-main}
\end{equation}
where $\gamma = 1.76 \times 10^{11}~\textrm{s}^{-1} \textrm{T}^{-1} $ is the gyromagnetic ratio and
$\alpha$ the dimensionless damping constant. Temperature is included
via Langevin dynamics by adding a Gaussian white noise $\boldsymbol{\xi}_i $
to the effective field 
$\boldsymbol{H}_{i}^{\rm eff} = -{\frac{\partial \mathcal{H} }{\partial{\boldsymbol{S}_i}}} + \boldsymbol{\xi}_i $. The two terms in the SLLG equation describe precession and damping dynamics through the torques acting on the spin system. 

In the literature, SOT is incorporated in the LLG equation as an additional torque term $\bm{T}_i^{\rm SOT}$ ~\cite{manchon2019}, 
\begin{equation}
     \boldsymbol{\dot{S}}_{i} = -\frac{\gamma}{\mu_{s}} \boldsymbol{S}_{i} \times \boldsymbol{H}^{\textrm{eff}}_{i} + \alpha \boldsymbol{S}_{i} \times  \boldsymbol{\dot{S}}_{i} + \frac{\gamma}{\mu_{s}}
     \boldsymbol{T}^{\textrm{SOT}}_{i} . \label{eq:llg-with-ext-sot}%
\end{equation} 
The form of the SOT can be derived from a quantum Hamiltonian analogous to the classical Hamiltonian Eq.~(\ref{eq:ham-with-field}), see {the} Supplemental Material, 
{Sec.\ III} \cite{SupMat}.
In general, $\boldsymbol{T}^{\textrm{SOT}}$ consists of both, damping-like and field-like torques, which in terms of a SOT field can be written as
\begin{equation}
   \boldsymbol{T}^{\textrm{SOT}}_{i}  = - \boldsymbol{S}_{i} \times  \boldsymbol{H}^{\textrm{SOT}}_{i}  ,\label{eq:ext-h_sot}
\end{equation}
where $\boldsymbol{H}^{\textrm{SOT}}_{i}$ is the field coming from the interaction of the strong moments with electrically-induced moments. In our model, these interactions are incorporated in the extended Heisenberg Hamiltonian,  Eq.~\eqref{eq:ham-renorm-with-field}, with $-\partial \mathcal{H}/\partial \boldsymbol{S}_{i}=\boldsymbol{H}^{\textrm{eff}}_{i}+\boldsymbol{H^{\textrm{SOT}}}_{i}$, where the first and third term in the Hamiltonian 
result in the effective field $\boldsymbol{H}^{\textrm{eff}}_{i}$ from strong moments and the remaining  terms collectively give $\boldsymbol{H^{\textrm{SOT}}}_{i}$ on differentiating the Hamiltonian with respect to the strong moments. From Eq.~\eqref{eq:ham-renorm-with-field}, it can be clearly seen that $\boldsymbol{H^{\textrm{SOT}}}_{i}$ is a linear function of the induced moments and $\boldsymbol{T}^{\textrm{SOT}}_{i}$ can, hence, be represented as
\begin{align}
   \boldsymbol{T}^{\textrm{SOT}}_{i} &   \propto  \boldsymbol{S}_{i} \times \left [ \delta\bm{\mu}_{S} \right]_{i/\nu}  
     = \boldsymbol{S}_{i} \times \left [  \delta\bm{\mu}_{S}^{e} +  \delta\bm{\mu}_{S}^{o}  \right]_{i/\nu}  ,\label{eq:ext-t_s-sot}
   \end{align}
for the induced spin moments in $\boldsymbol{H^{\textrm{SOT}}}_{i}$, and
\begin{align}
   \boldsymbol{T}^{\textrm{SOT}}_{i} &   \propto  \boldsymbol{S}_{i} \times \left [ \delta\bm{\mu}_{L} \right]_{i/\nu}  
     = \boldsymbol{S}_{i} \times \left [  \delta\bm{\mu}_{L}^{e} +  \delta\bm{\mu}_{L}^{o}  \right]_{i/\nu}  ,\label{eq:ext-t_l-sot}
   \end{align}
for the induced orbital moments in $\boldsymbol{H^{\textrm{SOT}}}_{i}$, where each term can be split in $M$-even and $M$-odd components.  
Taking, as an example, the spin part of $\boldsymbol{T^{\textrm{SOT}}}_{i}$,  each $\delta\bm{\mu}_{S}$ term can be expanded according to Eqs.~\eqref{eq:M-even_directions} and \eqref{eq:M-odd-directions}, so that 
$\boldsymbol{T^{\textrm{SOT}}}_{i}$ can be described as, 
    \begin{align}        \boldsymbol{T^{\textrm{SOT}}}_{i} & \propto  P(A) \,\boldsymbol{S}_{i} \times   (\boldsymbol{E} \times \hat{\boldsymbol{z}}) \nonumber \\
         & - P(B) \,\boldsymbol{S}_{i} \times  (\boldsymbol{S} \times (\boldsymbol{E} \times \hat{\boldsymbol{z}})) \nonumber \\
         & +P(A')\,(\boldsymbol{S}  \cdot \boldsymbol{E} )\boldsymbol{S}_{i} \times (\boldsymbol{S} \times \hat{\boldsymbol{z}})\nonumber \\ 
         & + P(B') \, (\boldsymbol{S} \cdot \boldsymbol{E}) \boldsymbol{S}_{i} \times \hat{\boldsymbol{z}} \nonumber \\
    & + P(A''') \, (\boldsymbol{S} \cdot \hat{\boldsymbol{z}})\boldsymbol{S}_{i} \times  (\boldsymbol{E} \times \boldsymbol{S}) .
    \label{eq:higher-order_tsot}
 \end{align}

Here, the $1^{\rm st}$ and $2^{\rm nd}$ term in Eq.~\eqref{eq:higher-order_tsot}  are the conventional field-like and damping-like forms of SOT, 
whereas in case of an in-plane magnetization such that $\boldsymbol{S} \, || \, \boldsymbol{E} $, the $3^{\rm rd}$ and $4^{\rm th}$ terms additionally act on the magnetization as planar field-like and planar damping-like torques. Also, there is another additional field-like torque term, the $5^{\rm th}$ term in Eq.~\eqref{eq:higher-order_tsot} that acts  in case of an out-of-plane magnetization aiming to switch it in the direction of the electric field, in a way 
that could also be seen as an initial stimulus for planar torques. Similar dependences apply to the orbital moments as well, where the torque caused by induced orbital moments $\delta \bm{\mu}_L$ acts via SOC  on $\bm{S}_i$ as described in Sec.~\ref{sec:III}. 

The dependence of the induced moments on the magnetization direction is incorporated in our model via the susceptibilities $\chi^{S/L} (\theta,\phi)$ where the angles describe the direction of the total magnetization for the Co induced moments and the direction of the interfacial magnetization for the Pt induced moments.
As the corresponding torques change with progressing  magnetization direction, the switching becomes a strongly nonlinear process. Nonetheless, and although the appearance of the $M$-even and $M$-odd induced moments is entangled, we can analyze their separate contributions by switching on or off their effects in the spin-dynamics simulations.

For the in-plane electric field in the $\hat{\bm{x}}$ direction, the induced spin moments can be separated into $M$-even and $M$-odd components as
\begin{align}
\frac{ \delta\bm{\mu}_{S}^e}{E_x}= \! \! \begin{bmatrix} [C_{2}]^{S}[1-\cos2\theta] \sin2\phi \\
	  -[C_{3}]^{S} -[C_{2}]^S [1-\cos2\theta]\cos2 \phi  + [C_{4}]^{S}\cos2\theta  \\
	[C_{8}]^{S}\sin2\theta \sin\phi
	\end{bmatrix} 
 \label{eq:M-even_moments}
 \end{align}
and  
\begin{align}
     \frac{\delta\bm{\mu}_{S}^o}{E_x}=\begin{bmatrix} [C_{1}]^{S}\cos\theta \\
	  0  \\
	[C_{7}]^{S}\sin \theta \cos \phi \\
	\end{bmatrix} ,
 \label{eq:M-odd_moments}
\end{align}
and similarly for the $M$-even and $M$-odd induced orbital moments. In addition, the influence of the induced orbital moments can be probed by including or neglecting them.

By transforming Eq.~\eqref{eq:llg-with-ext-sot} to the explicit form and implementing Eq.~\eqref{eq:ext-h_sot}, we obtain finally \cite{Selzer2022}
\begin{align}
      \boldsymbol{\dot{S}}_{i} & = -\frac{\gamma}{(1+\alpha^{2})\mu_{s}} \boldsymbol{S}_{i} \times \left[ \boldsymbol{H}^{\textrm{eff}}_{i} + \boldsymbol{H}^{\textrm{SOT}}_{i}\right] \nonumber \\
     & -\frac{\alpha \gamma}{(1+\alpha^{2})\mu_{s}} \boldsymbol{S}_{i}  \times \left(\boldsymbol{S} \times \left[\boldsymbol{H}^{\textrm{eff}}_{i} + \boldsymbol{H}^{\textrm{SOT}}_{i} \right]\right).
     \label{eq:llg-torque} 
\end{align}
Note that the torque term $\bm{T}_i^{\rm SOT}$ from Eq.\ (\ref{eq:llg-with-ext-sot}) now appears as an induced torque in both, precessional and damping terms.

These torque components are analyzed in detail in the following sections, with respect to their contributions from electrically-induced spin and orbital moments in Co and Pt. 

We performed simulations of the model described above for a system size of $128 \times 128\times 8$ Co spins with periodic boundary conditions in the lateral directions. The resulting magnetization and torque dynamics are analyzed layerwise for the 8 cobalt layers  {stacked} in the $\hat{\bm{z}}$-direction.

\section{Results}

\subsection{Switching behavior}

We investigate the magnetization dynamics of the ferromagnetic cobalt layer under the influence of electrically-induced spin and orbital moments, for a damping constant of $\alpha=0.05$~\cite{Barati2014} at room temperature   ($T=300$\,K), applying a continuous electric field in the 
$\hat{\bm{x}}$-direction.

According to our \textit{ab initio} calculations our model 
shows an easy-plane anisotropy. 
Therefore, the initial magnetization direction in the cobalt is set to the in-plane $\hat{\bm{x}}$-direction. 
We compare the magnetization dynamics 

in the presence or absence of induced orbital moments in addition to the induced spin moments for varying electric field magnitudes. As is evident

from Eq.~(\ref{eq:ind_mu_propto_E}), the induced moments and with that the SOTs are directly proportional to the electric field 
amplitude.

\begin{figure}
\includegraphics[scale=0.85]{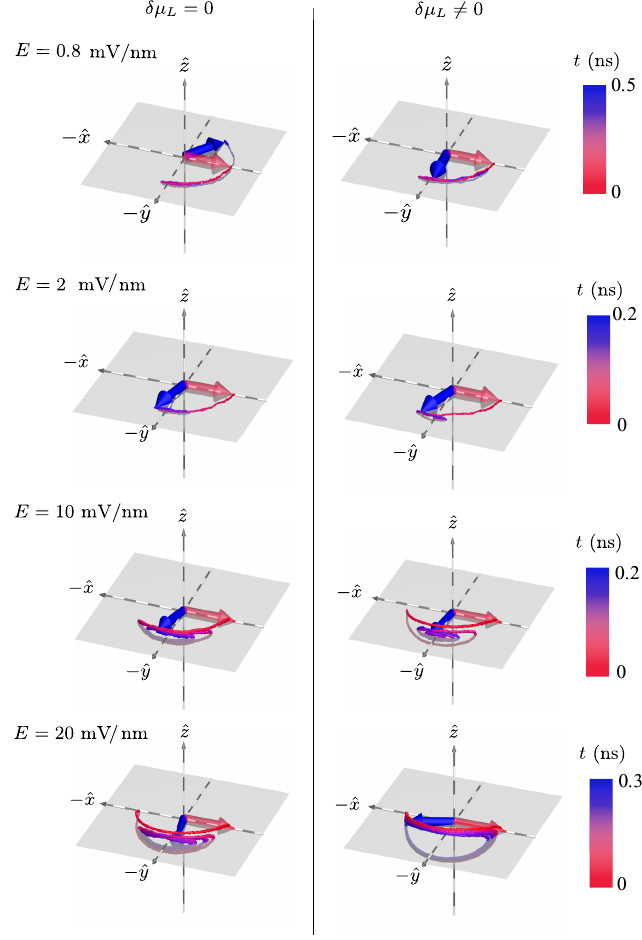}
\caption{Electric-field-induced magnetization switching of the Co magnetization in a Pt/Co bilayer. The figure compares simulations with ($\delta \mu_L \neq 0$) and without ($\delta \mu_L =0$) induced orbital moments for electric field values as indicated at room temperature.}
\label{fig:switching_at_low_field}
\end{figure}               
                
To start with, we 
study the 
dynamics of our model in the presence of all the induced moments and compare with the absence of induced orbital moments.  As can be seen in  
    Fig.~\ref{fig:switching_at_low_field}, in the presence of induced orbital moments ($\delta \mu_{L} \neq 0$) the $90$-degree in-plane switching begins to appear  at comparatively lower electric field magnitude ($E=0.8$ \text{ mV/nm}), where the switching direction is given by $\boldsymbol{E} \times \hat{\boldsymbol{z}}$. The SOT switches the magnetization from the $\hat{\boldsymbol{x}}$-direction to the $-\hat{\boldsymbol{y}}$-direction. As the switching direction is in-plane and since the system has easy-plane anisotropy only, i.\ e., no anisotropy energy barrier in plane, switching occurs when SOT effects overcome the thermal fluctuations,  There is, however, no final steady state as long as the  SOT induced field is weaker than the thermal fluctuations and the magnetization vector will fluctuate in the easy plane as can be seen at low electric field ($E=0.8$ mV/nm) and in absence of induced orbital moments ($\delta \mu_{L}=0$).
The influence of the induced spin moments alone ($\delta\mu_L =0$) is much weaker as now the switching can be observed only 
for higher electric fields ($E \geq 2$ {mV/nm}). 

\begin{figure}               
    \includegraphics[scale=0.85]{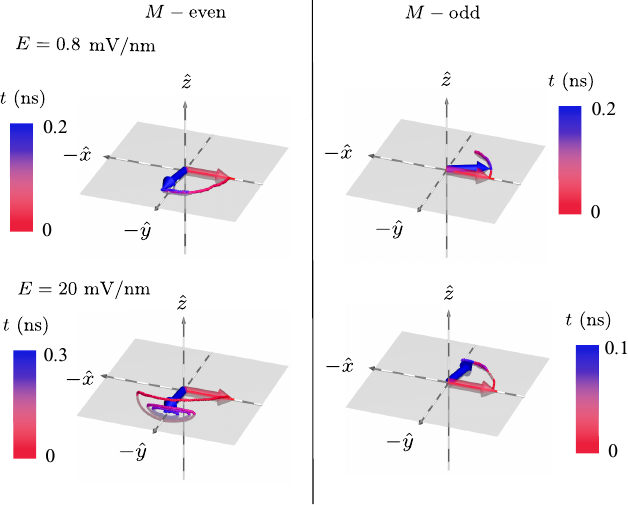}
    \caption{Simulated switching of the Co magnetization in the Pt/Co bilayer, when only  $M$-even (left), and $M$-odd (right) induced moments are taken into account. 
    }   
    \label{fig:non-magnetic_ev_od_moment_ind_switch}
\end{figure} 

On further increasing the electric-field strength it  can be observed that  
the system first exhibits faster switching  ($E = 2$ {mV/nm} and $10$ {mV/nm}), and then begins to 
oscillate at even
higher fields again not reaching a final steady state. 
Finally, for fields of $E = 2$ {mV/nm} and $10$ {mV/nm} the system switches  in $-\hat{\boldsymbol{y}}$ direction for  both, presence and absence of induced orbital moments, but  with differences in switching and relaxation times. E.g., for $E=10$ mV/nm the magnetization has reached its final state only in the presence of induced orbital moments, while the magnetization is still precessing on the same time scale in  the absence of induced orbital moments.

To understand the peculiar oscillating behavior for higher fields, we have separated the influence of $M$-even and $M$-odd dependent induced moments. As expressed in Eqs.\ (\ref{eq:M-even_directions})  and (\ref{eq:M-odd-directions}), the induced moments have magnetization-independent ($P(A)$ term) and magnetization-dependent { $M$-even (terms with $P(A')$, $P(A'')$ and $P(A''')$) and $M$-odd (terms with $P(B)$ and $P(B')$) origins. The influence of  $M$-even  and $M$-odd induced moments can nonetheless be separately studied. }

As depicted in Fig.~\ref{fig:non-magnetic_ev_od_moment_ind_switch}, it is the $M$-even induced moments that support switching in the $\boldsymbol{E} \times \hat{\boldsymbol{z}}$ direction at any electric field magnitude, whereas the $M$-odd moments are weaker in magnitude and  begin to contribute significantly only at higher electric fields (here, $E=20$ {mV/nm}). Moreover, these $M$-odd moments support a switching direction given by $-\boldsymbol{S} \times (\boldsymbol{E} \times \hat{\boldsymbol{z}})$ as well as  $(\boldsymbol{S} \cdot \boldsymbol{E}) \boldsymbol{\hat{z}}$, which thus applies  torque in the out-of-plane direction, facing limits set by the anisotropy of the system and applies an anti-damping torque that moves the magnetization towards $+\hat{\boldsymbol{y}}$ direction, if stronger than the thermal energy (at higher field $E=20$ {mV/nm}), as we will see in the torque section in detail.

This competition between the preferred directions of switching  explains why we observed oscillatory dynamics in our case of in-plane anisotropy at higher electric fields when both $M$-even and $M$-odd moments become strong enough to influence the magnetization dynamics of the system.

The time scale of the switching process can be quantified as the time it takes for the magnetization to reach $-\bm{\hat{y}}$  with $M_{x}=0$  for the first time after having applied the field. For the cases discussed above, it varies from $90$ ps (at low electric field)  to  $13$ ps (at $E=10 $ mV/nm) for magnetization-dependent moments when the induced orbital moments are present, whereas slower rates are obtained in the absence of induced orbital moments, as can be expected (see Sec.\ V in Supplemental Material \cite{SupMat}). 

This is in line with results shown in Fig.\ \ref{fig:MES_constants} where the average contribution from the  $M$-even spin moments over the layers of Co is smaller than its orbital counterpart.

In the case of magnetization-independent moments, the switching time is almost 
the same 
with or without induced orbital moments and faster than for the magnetization dependent case. Moreover, switching occurs  with only magnetization-independent spin moments for lower electric fields than in the presence of magnetization-dependent spin moments where it does not switch (see Sec.\ V in Supplemental Material \cite{SupMat}). 
This means that the magnetization-dependent induced moments are crucial to explain the effect of induced orbital moments on the switching, as will be discussed in the next subsection in connection with the layer-resolved torques.

The orbital moments also enhance the switching speed in simulations performed at $T=0$ K, as shown in 
Sec.\ VI in the Supplemental Material~\cite{SupMat}.

From the above observations and the symmetry analysis of the induced moments in Eqs.\ \eqref{eq:M-even_directions} and \eqref{eq:M-odd-directions} as well as Fig.\ \ref{fig:MES_constants}, it can be concluded that the presence of  induced orbital moments

contribute largely to the magnetization switching, as  these point in $\boldsymbol{E} \times \hat{\boldsymbol{z}}$ direction and are dominant, which can be concluded  from the size of  $P(A)$ and $M$-even $P(A') $ and $P(A''')$ for induced orbital moments
in Fig.\ \ref{fig:MES_constants}. 

The $M$-even magnetization-dependent spin and orbital moments  in Eq.~\eqref{eq:M-even_directions} support switching in the in-plane direction,

whereas $M$-odd induced moments with coefficient $P(B)$ and $ P(B')$ would contribute to out-of-plane switching.

Hence, the $M$-even induced moments are useful for SOT switching of magnetic layers with in-plane anisotropy, whereas the $M$-odd moments are useful for systems with out-of-plane magnetic anisotropy \cite{Lee2020}. The latter type of switching has been observed, too, experimentally \cite{MihaiMiron2010,manchon2019}. 
 
To delve deeper into the underlying torque mechanisms and gain further information, we have studied the influence of the interaction terms in Eq.\ (\ref{eq:ham-renorm-with-field}). In the following subsection we explore the field-like and damping-like behavior of the SOT arising from the different origins of the induced moments.

 \subsection{Layer-resolved spin and orbital torque }

In the following we conduct a  layer-resolved analysis of the 
SOT arising from induced spin and orbital moments 

by dissecting the contributions of each term to the effective field $\boldsymbol{H}_{i}^\text{SOT}$ in Eq.\ (\ref{eq:ext-h_sot}). Since our simulations demonstrate that the system switches towards the $\boldsymbol{P}=\boldsymbol{E} \times \hat{\boldsymbol{z}}$ direction when all torque components are taken into account, we separate these torque terms according to Eq. \eqref{eq:llg-torque} into field-like contributions, causing a precession of the spin $\boldsymbol{S}_{i}$ around $\boldsymbol{P}$, and damping-like contributions,  acting in the plane spanned by $\boldsymbol{S}_{i}$ and $\boldsymbol{P}$ as: 
\begin{align}
    \bm{\mathcal{T}}_i^\text{SOT}   
    & =   \, \bm{T}_{i}^\text{SOT} + \alpha  \boldsymbol{S}_{i} \times  \bm{T}_{i}^\text{SOT} \nonumber \\
    & =  -\boldsymbol{S}_{i} \times \left( T_{\text{F}}  \left(  \boldsymbol{P}\right) + T_{\text{D}}  \left( -
    \boldsymbol{S} \times \boldsymbol{P}\right)\right) \nonumber \\
   & ~~~ -\alpha \boldsymbol{S}_{i} \times  \left( \boldsymbol{S}_{i} \times \left(  T_{\text{F}}  \left( \boldsymbol{P} \right) +T_{\text{D}}  \left( -\boldsymbol{S} \times \boldsymbol{P}\right)\right)\right).
   \label{eq:FL_Dl_torque_llg}
   \end{align}
   \begin{figure*}[t]     
      \includegraphics[scale=0.9]{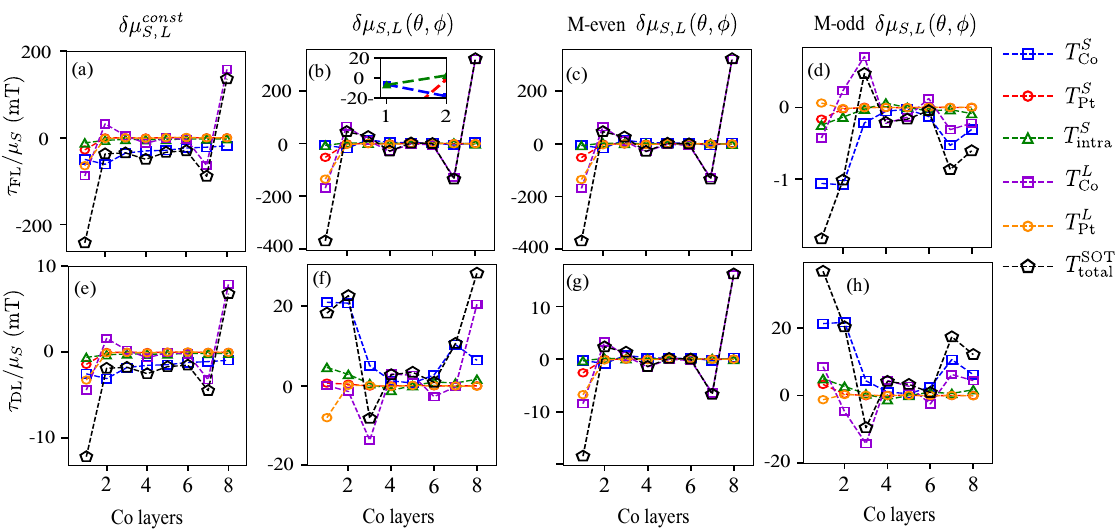}
    
    \caption{Magnitude of the layer-resolved initial field-like and damping-like SOT components for $E =0.8$ mV/nm, separating the contributions from magnetization-independent, magnetization-dependent, $M$-even and $M$-odd induced moments $\delta \mu_{S,L}$, respectively. 
Layer 1 is the interface layer to the Pt, layer 8 is the surface layer of the Co film. 
    The colored symbols
   depict different contributions to the total SOT ($T_{\text{total}}^{\text{SOT}}$), including the torques due to the interaction of local moments with induced spin moments in Co via exchange $(T^{S}_{\text{Co}})$, via intra-atomic exchange $(T^{S}_{\text{intra}})$, with induced spin moments at the interfacial Pt via renormalized exchange $(T_{\text{Pt}}^{S})$, with the induced orbital moments in Co via spin-orbit coupling $(T_{\text{Co}}^{L})$, and with induced orbital moments at the Pt interface via renormalized spin-orbit interaction $(T_{\text{Pt}}^{L})$, respectively. 
   }
    \label{fig:tq_layer_2e7}
\end{figure*}
According to Eq.~\eqref{eq:higher-order_tsot}, $T_{\text{F}}$ has contributions from terms with coefficients $P(A)$, $P(A')$ and $P(A''')$ and $T_{\text{D}}$ from terms with coefficient $P(B)$ and $P(B')$.

   For any atomic layer of Co, the above torques can be averaged and expressed as 
\begin{align}
      \bm{\mathcal{T}}^\text{SOT}   = &- (T_{\text{F}} + \alpha T_{\text{D}}) \boldsymbol{S} \times \boldsymbol{P} \nonumber \\
     & - (\alpha T_{\text{F}} -T_{\text{D}} )\, \boldsymbol{S} \times (\boldsymbol{S} \times \boldsymbol{P} ) \nonumber \\
      = &  \, \tau_{\text{FL}}  \boldsymbol{\hat{u}}_{S \times P}+ \tau_{\text{DL}} \boldsymbol{\hat{u}}_{S \times \left(S \times P\right)}\, .
     \label{eq:FINAL_FL_DL}
\end{align}

Here, $\tau_{\text{FL}}$ and $\tau_{\text{DL}}$ 
quantify the field-like and damping-like contributions to the total SOT. 

For a realistic $E$-field value for switching, we here focus on lower electric field magnitude, $E=0.8$ {mV/nm}.
When considering the total induced moments, the small damping constant $\alpha$ results in $T_{\text{F}}$  dominating the $\tau_{\text{FL}}$ term and $T_{\text{D}}$ dominating the $\tau_{\text{DL}}$ term of Eq.~(\ref{eq:FINAL_FL_DL}). Also, for the case of magnetization-independent moments, there will be only the $T_{\text{F}}$ term in the net field-like and  damping-like torques of the LLG equation.

The layer-resolved initial torque components $\tau_{\text{FL}}$ and $\tau_{\text{DL}}$ are shown in Fig.~\ref{fig:tq_layer_2e7},  separated according to the presence of various current-induced moments which are considered in our simulations. 
Here, for a switching direction given by $\boldsymbol{P}$, 

an initial damping-like torque acting in direction of $\boldsymbol{P}$, is referred to as a damping torque whereas the torque acting in $-\boldsymbol{P}$ direction is referred to as being anti-damping.

First, comparing the total ($T_{\text {total}}^\text {SOT}$) net field-like and damping-like torques for the cases of  magnetization-independent induced moments ($\delta\mu_{S,L}^{\text{const}}$, Fig.~\ref{fig:tq_layer_2e7}(a) and (e))  and in the presence of additional magnetization-dependent induced moments ($\delta \mu_{S,L} (\theta,\phi)$, Fig.~\ref{fig:tq_layer_2e7}(b) and (f)),  it is evident that the field-like SOT

is larger than its damping-like counterpart. 

Second, focusing on $\tau_{\text {FL}}$ in these two cases ($\delta\mu_{S,L}^{\text{const}}$ and $\delta \mu_{S,L} (\theta,\phi)$), we see that the contributions from interface and surface layers of Co are dominating,

where the larger contribution from the interface layer signifies that switching is driven by the interface 
SOT for both these cases. Furthermore, on comparing the spin and orbital contributions at the interface layer, it is observed that in both cases 
the largest contribution comes from the orbital torque due to the orbital moments induced in Co via spin-orbit coupling ($T_{\text {Co}}^L$), and the second largest contribution is the torque from the  orbital moments at the Pt interface $T_{\text {Pt}}^L$ 
via renormalization. 
When we compare the two columns ($\delta\mu_{S,L}^{\text{const}}$ and $\delta \mu_{S,L} (\theta,\phi)$) for $T_{\text {total}}^\text {SOT}$ and the other contributions, we see that the overall field-like torque is enhanced in the presence of magnetization-dependent moments due to the enhancement of orbital torque contributions ($T_{\text{ Pt}}^L$ and $T_{\text {Co}}^L$).  
We also observe a reduction of the  net contribution from the induced spin moments in Co ($T_{\text {Co}}^S$ and $T_{\text {intra}}^S$) in the presence of additional  $M$-dependent induced moments,  meaning that the latter balance the opposite $M$-independent spin counterpart.
Thus, the strongest contribution supporting switching is coming from the orbital torque at the interface layer, through spin-orbit interactions in Co and renormalized interactions at the Pt interface layer. 
The weaker contribution of the spin torque $T_{\rm Pt}^S$, $T_{\rm Co}^S$ and $T_{\rm intra}^S$ substantiates why switching does not occur in the absence of induced orbital moments at low electric fields. {This reduction in spin contribution at {the} interface has also been observed in previous works and has been discussed in terms of interface and proximity effects in {N}M/FM systems \cite{Amin_2016,Dolui2017,Zhu2019}.}

Note that the induced Co orbital moments ($T_{\text{Co}}^{L}$) have the strongest contribution in the surface layer and their field-like torque component is opposite to the torque in the interfacial Co layer. This indicates that 
the interaction from the interfacial Pt layer which is taken into account via renormalization ($T_{\text{Pt}}^{L}$ and $T_{\text{Pt}}^{S}$) plays a vital role in determining the switching direction. These renormalization terms not only support switching in the same direction but also, being additional terms, they counter the opposite switching effects coming from the induced orbital moments in the surface layers of Co {(see details in Sec.\ VII of {the} Supplemental Material \cite{SupMat}).}

Third, considering the net damping-like torque ($\tau_{\text{DL}}$,  {bottom row) in Fig.~\ref{fig:tq_layer_2e7}} in the two cases of magnetization independent and additional magnetization-dependent SOTs, 
we note that the interface contribution is higher 
in the absence of magnetization-dependent moments and the initial damping-like torque at the interface is purely dissipative
, whereas in the presence of magnetization-dependent moments the total SOT initially has an anti-damping character from both, interface and surface layers (note the positive sign). This can be understood from  Eq.\ (\ref{eq:FINAL_FL_DL}) since  for $M$-independent moments there exists only the $\alpha T_{\text F}$ term while for $M$-dependent moments there will be additional $M$-even terms in $\alpha T_{\text F}$ and $M$-odd terms in $ T_{\text D}$. 
Focusing on the interface layer only, it can be observed that the SOT due to the exchange field from induced spin moments in Co ($T_{\text{Co}}^{S}$) possesses strong anti-damping character, whereas 
the orbital torque from induced orbital moments at the Pt interface  ($T_{\text{Pt}}^{L}$)  has a damping character, 
and the orbital torque from induced orbital moments in Co ($T_{\text{Co}}^{L}$) has a negligibly small contribution to the damping-like torque. 

Fourth, we focus on the role played by the  
$M$-even and $M$-odd induced moments. On extracting the torque from  $M$-even induced moments, it can be seen in Fig.~\ref{fig:tq_layer_2e7}(c) and (g) that the SOT 
 possesses a strong field-like character and the weaker $M$-even damping-like torque, smaller by a factor of $\alpha$, supports switching in the $\boldsymbol{E} \times \hat{\boldsymbol{z}}$ direction. 
 Additionally, at the interface layer there is no sign of an $M$-even contribution to the anti-damping torque, whereas the torque from $M$-odd induced moments (Fig.~\ref{fig:tq_layer_2e7}(h)) possesses strong anti-damping-like character. This confirms that the anti-damping character visible in presence of all magnetization-dependent moments ($\delta \mu_{S,L} (\theta,\phi)$) at the interface layer is due to $M$-odd moments. This anti-damping torque at the interface layer 
affects the magnetization relaxation rate and can lead to a stationary non-equilibrium state if it is strong enough to compete with the torque from the $M$-even moments, an effect which we have shown for higher electric field values in Fig.\ \ref{fig:switching_at_low_field}.  

The information regarding the initial torques that start the switching dynamics  has been sufficient to study the roles of various torque contributions at different layers of the Co film. However, for a deeper understanding of the role of  magnetization-dependent induced moments, especially the presence of anti-damping torques, it is needed to study the torque dynamics.

\begin{figure}
      \centering
 \includegraphics[scale=0.9]{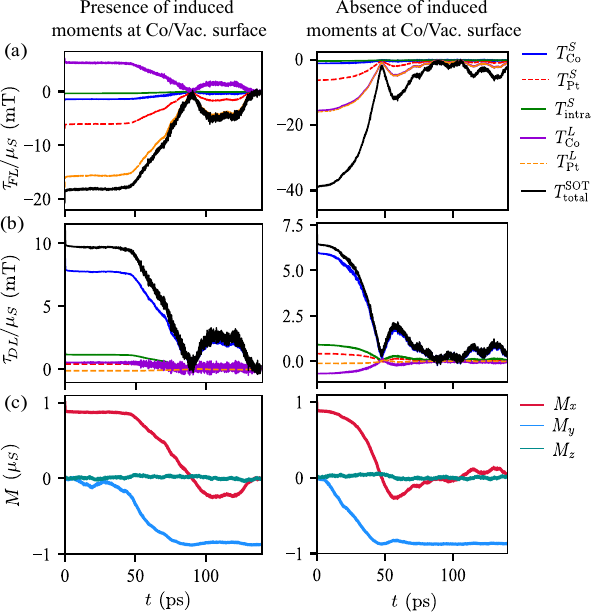}

   \caption{Dynamics of the average field-like (a) and damping-like (b) torque acting on the Co layer, separated according to different SOT contributions as specified in the legend for an  electric field value of $E=0.8$ mV/nm in {the} presence (left {column}) and absence (right {column}) of induced moments at Co/vacuum surface. {Panel} (c) shows the corresponding {simulated} magnetization dynamics.}
\label{fig:Average_tq_dynamics_8e5}
\end{figure} 

To this end we have extracted the dynamics of the layer-averaged field-like and damping-like torques $\tau_{\rm FL}$ and $\tau_{\rm DL}$ acting on the cobalt film.  As shown in Fig.~\ref{fig:Average_tq_dynamics_8e5} (left column)
we have first analyzed the torque dynamics leading to magnetization switching in the presence of induced moments at the Co/vacuum surface, where the magnetization dynamics [Fig.~\ref{fig:Average_tq_dynamics_8e5}(c)] follows the switching path $\boldsymbol{E} \times \hat{\boldsymbol{z}}$ and there is no influence of anti-damping torque $\tau_{\rm DL}$. This further confirms that $\tau_{\rm DL}$ arising from $M$-odd moments is inefficient and switching is solely driven by field-like torque from $M$-even moments. In this case a large contribution is coming from orbital torque at the Pt interface ($T^{L}_{\rm Pt}$). Since we have seen in Fig.\ \ref{fig:tq_layer_2e7} that the Co/vacuum surface layer applies torque in the opposite direction to that of the interface, and thereby reduces the overall effect of SOT, 
we have also simulated and compared the torque and magnetization dynamics in the absence of those induced moments at the Co/vacuum surface. This is highly important, since in experiments usually capping layers are used to diminish surface effects.

As shown in Fig.~\ref{fig:Average_tq_dynamics_8e5} (right), in this case the field-like torque has strong contributions from the orbital torque arising from both, the interfacial Pt and Co layers, thus enhancing the overall field-like torque compared to the case which induces the effects of the Co/vacuum surface layer. This

leads to faster switching with a switching time of $45$ ps, as can be seen from magnetization dynamics, whereas switching observed in the presence of Co/vacuum surface moments takes  $90$ ps. This {furthermore} indicates that in absence of {current-induced} Co/vacuum surface moments the switching can occur at even lower electric field {strengths}. {Specifically,} we have observed {switching} at {a} minimum field of $E_{x}=0.4 $ mV/nm with a switching time of $85$ ps, described in detail in Sec.\ VII of  the Supplemental Material \cite{SupMat}.
{Lastly,} we have also analyzed the SOT-field dynamics for the SOT-interaction terms in Eq.\ \eqref{eq:ham-renorm-with-field} 

and observed that {the SOT field} from the $M$-even induced spin moments is strongly magnetization dependent whereas the {SOT field} from $M$-even orbital moment interaction has negligibly small variations with {change{s of the}  in-plane} magnetization directions, see Sec.\ VIII of {the} Supplemental Material \cite{SupMat}.  This explains {why our simulations uncover} 
{a} large role played by the 
{$M$-even} 

field-like

interface orbital torque in driving the {90-degree} magnetization switching {of the} in-plane anisotropic Pt/Co {system}.

\section{Discussion and Conclusions}

Our {layer-resolved} microscopic model for the NM/FM system Pt/Co facilitates a quantitative examination of the different origins and mechanisms of spin-orbit torque. Magnetization-independent and magnetization-dependent induced moments have been separated using  magneto-electric susceptibility tensors calculated from first principles, which, based on their distinct dependence on $M$, $E$, and the proximity to the interface, revealed the strength of induced spin and orbital moments. Orbital moments induced due to the nonrelativistic OHE and OREE in Co and at the Pt/Co interface are comparatively stronger whereas the orbital moments deeper in the Pt  
layer are significantly weaker. Our analysis reveals the {larger}
magnitude of $M$-even induced moments and weaker $M$-odd moments, and demonstrates  that both of these magnetization-dependent moments  are strongest at the {Pt/Co} interface.
Studying the magnetization switching under the action of the SOT, it could be clearly verified that the previously often disregarded induced orbital moments play {in fact} a vital role as these facilitate switching at lower electric field strengths {($0.8$ mV/nm, which is around $10^{8}$ A$\text{cm}^{-2}$ in terms of current density) with {sub-nanosecond} switching time of $90$ ps}. Our simulat{ed} results {have} revealed that {the} $M$-even induced moments, and in particular {the} orbital moments, support in-plane switching, whereas $M$-odd induced moments, specifically, {the} induced spin moments, drive the magnetization out of plane. 

{The spin part of the SOT has already been widely investigated (see \cite{manchon2019})  often in a macrospin approximation, yet not in a layer-resolved manner. Also, not much is known about the orbital contribution to the SOT and how it operates.  Hence, our atomic-layer and spin- and orbital-resolved approach provides additional insight. Orbital torques have been proposed in NM/FM bilayers \cite{Lee2021,Wang2024, Zhou2025, Krishnia2024, Wang2024_review}, but in experiments, such as spin-torque ferromagnetic resonance measurements, it is often difficult to separate spin and orbital torque contributions. Recently, Wang \textit{et al.}\ \cite{Wang2024} deduced a large damping-like torque due to the OREE in Pt/CoO bilayers, and Krishnia \textit{et al.}\ \cite{Krishnia2024} found a large damping-like torque in the Pt/Co/Cu/CuO system due to the OREE. In a different heterostructure, Pt/Co/Al/Pt, \mbox{Krishnia} \textit{et al.}\ \cite{Krishnia2023} found a large enhancement of the field-like SOT due to orbital texture and OREE at the Co/Al interface \cite{Nikolaev2024}. Consistent with this,  we find in our study a large field-like torque due to accumulation of orbital moments at the Pt/Co interface, even though in general, the SOT appears to depend on the specifics of the heterostructure.}
{Our simulated results are furthermore  in agreement with a recent experimental study of switching in Pt/CoFeB by Zhang \textit{et al.}\ \cite{Zhang2025}, who found that field-like SOT originating from the ferromagnetic layer can dominate the switching  process. The thus-induced switching could be fast ($\sim 0.2$ ns) and occur at relatively low electric field strengths \cite{Zhang2025}, consistent with our findings.}

{As} {our \textit{ab initio}-based simulations} stress the large contribution of the orbital torque {to the switching} it {deserves to} be {further analyzed in comparison to}

orbital-related torques in NM/FM bilayers {that} have drawn attention quite recently
\cite{Go2020,Go2020b,Lee2021,Go2023,Jo2024}.
It remains, however, a point of discussion how these {different} orbital torques arise. Go and Lee \cite{Go2020b} define an orbital torque due to orbital diffusion into the FM layer, where it is converted by spin-orbit coupling to a non-equilibrium spin polarization that couples to the local FM moments due to exchange interaction. 
It is possible that our computed spin torques $T_{\rm Co}^S$ in Co contain a contribution stemming from such a mechanism, but it does not appear as a dominant contribution.
Another recent work {by Go \textit{et al.}} defines the spin-orbital torque \cite{Go2023}, wherein the induced orbital angular moments couple through spin-orbit coupling to the local moments in the FM layer. This orbital-to-spin conversion mechanism tallies with our torques $T_{\rm Co}^L$ and $T_{\rm Pt}^L$ that are computed to be the largest ones. A distinction is, however, that in Ref.\ \cite{Go2023} this spin-orbital torque is predicted, for Cr/CoFe, to be large in the damping-like torque, whereas we find it conversely to be large in the field-like torque of Pt/Co. In addition, we find this spin-orbital torque to be rather short ranged, whereas it is long ranged in Ref.\ \cite{Go2023}, which suggests that it is {actually} a different type of orbital torque.

Since our analysis provides 
layer-resolved spin-orbit torques, it can be concluded that the torque  applied through induced spin moments at the interface layer by exchange interaction has a major contribution to an anti-damping behavior, arising from {the} $M$-odd induced spin moments. 
The {in-plane} switching occurs due to the action of a field-like SOT applied primarily via {the} $M$-even induced orbital moments at the Pt/Co interface.  From earlier work on current-induced orbital moments
in metallic bilayers \cite{salemi2021}, it is known that these $M$-even induced orbital
moments are independent of the strength of spin-orbit interaction and are thus of nonrelativistic origin.
However, they can apply a torque on the local spin moments only through the local spin-orbit
interaction. Thus, although being of nonrelativistic nature the induced orbital moments require spin-orbit interaction with local spin moments to apply a torque. 
The spin-orbit interaction from the Pt interface layer is taken into account via renormalization  
{and is confirmed} to be an important term supporting switching. 

We note that the net field-like torque for Pt/Co is larger than the (anti-)damping-like torque, in contrast to most previous Pt/Co experimental studies that consider comparatively thicker Pt layer. From our  first-principles calculations it is found that the $M$-even induced spin moments 
(pointing in the $\boldsymbol{E} \times \hat{\boldsymbol{z}} $ direction) have contribution arising mostly from the interface, which in turn applies a field-like torque, and the $M$-odd induced spin moments (pointing in $- \boldsymbol{M} \times \boldsymbol{E} \times \hat{\boldsymbol{z}}$ direction) have a mixed bulk and interface origin and apply an anti-damping like torque. Indeed, for the spin part, due to filtering and scattering effects at the interface, the net field-like contribution at the Co layers diminishes and the anti-damping torque from the spin part gets stronger. This is usually the case for thick Pt layers where the bulk SHE contribution dominates and generates stronger damping-like torque. In experiments \cite{Gabor2024}, it has indeed been shown that on decreasing the Pt thickness the bulk SHE contribution decreases leading to a reduced in damping-like effective-field. Moreover, it is difficult to clearly separate the effective field-like and damping-like fields for smaller Pt thickness where interface effects begin to dominate and a net increment in field-like torque begins.    
In our calculations it has also been shown 
that the strong orbital contribution comes from the interface effects which again explains that strong-field-like torque clearly dominates in a thin Pt/Co bilayer. Previously, 
it has been found \cite{Lee2021} that the dominating torques from induced orbital moments and induced spin moments have the same direction. In our case, these similar directions can be explained from the direction of the net effective field-like torque from $M$-even induced spin moments at the Co and Pt interface layer and $M$-even induced orbital moments, leading to switching in the same direction.

Additionally, it should be noted that the effect of spin diffusion is already present in the magnetoelectric susceptibility tensor computed via relativistic first-principles approach. However, it might indeed be a question whether the influence of spin diffusion would not be more pronounced when one considers a thick Pt layer, which in turn might enhance the effect of damping-like torque. 

{Our 
{simulations further} reveal 
that at elevated electric-field magnitude the  induced $M$-odd spin moments, {due to the magnetic spin Hall and {magnetic} Rashba-Edelstein 
effects}, with its anti-damping nature begin to interfere in the switching process. This anti-damping effect can be diminished by enhancing the Gilbert damping parameter for such system as can be recognized from Eq.~\eqref{eq:FINAL_FL_DL}. A lower damping constant leads to a stronger anti-damping torque in the system. For future experiments, it might be possible to reduce the strength of {$M$-odd} induced spin moments in Co 

by a 
{suitable} interface engineering. 
Moreover, we have observed that on turning off the {current-}induced moments at {the} Co/vacuum surface, the switching occurs {on a} faster time-scale, ($45$ ps at $E_{x}=0.8$~mV/nm) and can occur at {even} lower electric fields of magnitude $E_{x}=0.4$ mV/nm with switching time of $85$ ps. {Finally,}

we conclude that it is {the  large} 
$M$-even 

orbital moments from {the} Pt and Co interface layer{s} that drive the switching via {a} field-like torque at low electric field magnitude for in-plane anisotropic Pt/Co.}

Our quantitative treatment thus offers detailed insights into the required conditions  needed to 
achieve fast and deterministic switching at room temperature, and contributes to the distinctive understanding of the interplay between damping-like and field-like torques caused by induced spin and orbital moments from magnetization-independent and  interface-based magnetization dependent origins. \\


\begin{acknowledgments}
{We thank Daegeun Jo for valuable discussions.} This work has been supported by the the German Research Foundation (Deutsche Forschungsgemeinschaft) through CRC/TRR 227 (project MF, project-ID: 328545488)) and CRC 1432 (project B02, project-ID: 425217212), the Swedish Research Council (VR), the K.\ and A.\ Wallenberg Foundation (Grants No.\ 2022.0079 and 2023.0336),  by the EIC Pathfinder OPEN grant No.\ 101129641 ``OBELIX'', by the National Research, Development, and Innovation Office (NRDI) of Hungary under Project Nos.\ K131938, K142652, FK142601 and ADVANCED 149745, by the Thematic Area Excellence Program of the Ministry of Culture and Innovation from the NRDI Fund through the Grant No.\ TKP2021-NVA-02, and by the Hungarian Academy of Sciences via a J\'{a}nos Bolyai Research Grant (Grant No.\ BO/00178/23/11). Computational resources were provided by the core facility SCCKN in Konstanz and the National Academic Infrastructure for Supercomputing in Sweden (NAISS) at NSC 
Link\"oping, partially funded by VR through Grant Agreement No.\ 2022-06725.
\end{acknowledgments}

\bibliography{SOT_Pt-Co_manuscript}

\end{document}